\documentclass[%
reprint,
amsmath,amssymb,
aps,
pra,
floatfix,
]{revtex4-1}

\usepackage{color}
\definecolor{coolblack}{rgb}{0.0, 0.18, 0.39}

\usepackage{graphicx}
\graphicspath{ {figures/} }
\usepackage{dcolumn}
\usepackage{bm}
\usepackage[linktocpage,colorlinks = true,linkcolor = blue,urlcolor  = coolblack,citecolor = red,anchorcolor = green]{hyperref}
\usepackage[utf8]{inputenc}

\begin{document}


\pdfstringdefDisableCommands{\def\\{}}

\title{Gravitational waves from scale-invariant vector dark matter model:\\
Probing below the neutrino-floor}

\author{Ahmad Mohamadnejad}
\email{mohamadnejad.a@lu.ac.ir}
\affiliation{Department of Physics, Lorestan University, khorramabad, Iran}
\affiliation{School of Particles and Accelerators, Institute for Research in Fundamental Sciences (IPM), Tehran, Iran}

\date{\today}

\begin{abstract}
We study the gravitational waves (GWs) spectrum produced during the electroweak phase transition in a scale-invariant extension of the Standard Model (SM), enlarged by a dark $ U(1)_{D} $ gauge symmetry. This symmetry incorporates a vector dark matter (DM) candidate and a scalar field (scalon). Because of scale invariance, the model has only two independent parameters and for the parameter space constrained by DM relic density, strongly first-order electroweak phase transition can take place. In this model, for a narrow part of the parameter space, DM-nucleon cross section is below the neutrino-floor limit, and therefore, it cannot be probed by the future direct detection experiments. However, for a benchmark point from this narrow region, we show the amplitude and frequency of phase transition GW spectrum fall within the observational window of space-based GW detectors such as eLISA.
\end{abstract}

\maketitle


\section{Introduction \label{sec1}}

The detection of GWs \cite{Abbott:2016blz} has opened up a new and independent avenue for probing of dark matter \cite{Bertone:2019irm}. These waves are ripples in the fabric of space-time generated by energetic and violent sources such as black hole and neutron star binaries, extreme mass ratio inspirals, and first order cosmological phase transitions.
The main targets of ground-based GW detectors are black hole and neutron star binaries with best sensitivity at frequencies
$ {\cal{O}} (10^2) $ Hertz, while space-based detectors are most sensitive to milli-Hertz or deci-Hertz frequencies \cite{Pitkin:2011yk}. The stochastic background of primordial GWs produced during first order electroweak phase transition is a physical sources of GWs in this frequency band \cite{Caprini:2015zlo}.

Cosmic phase transitions occur when the temperature drops below a critical temperature leading to the transition of the Universe from a symmetric phase to a phase of broken symmetry (for a recent review see \cite{Mazumdar:2018dfl}). 
In the SM, electroweak phase transition, as well as QCD phase transition, is of second order \cite{Kajantie:1996mn,Aoki:1999fi} and does not generate the GW signal.
However, early Universe might be in a state where not only the gauge symmetry was present but also a (classical) scale-invariant or conformal symmetry was realized, preventing any massive parameters in the Lagrangian.
Then the quantum effects must have broken the scale-invariant symmetry, generating a nonzero VEV of the scalar field(s) and all the mass terms of massive particles via Coleman-Weinberg mechanism \cite{Coleman:1973jx}.
Particularly, conformal DM models with Higgs portal are attractive because they can solve DM problem and at the same time can generate a strongly first order phase transition \cite{Farzinnia:2014yqa,Sannino:2015wka,Ghorbani:2017lyk,YaserAyazi:2019caf}.
GWs due to first order phase transition have been studied within models where the scale-invariant symmetry
is broken due to Coleman-Weinberg mechanism \cite{Espinosa:2008kw,Dorsch:2014qpa,Jaeckel:2016jlh,Hashino:2016rvx,Jinno:2016knw,Marzola:2017jzl,Brdar:2018num,Prokopec:2018tnq,Marzo:2018nov} or models with DM candidate \cite{Schwaller:2015tja,Dev:2016feu,Chala:2016ykx,Baldes:2017rcu,Chao:2017vrq,Beniwal:2017eik,Huang:2017rzf,Huang:2017kzu,Addazi:2017gpt,Addazi:2017nmg,Hashino:2018zsi,Croon:2018erz,Bian:2018mkl,Bian:2018bxr,Shajiee:2018jdq,Madge:2018gfl,Baldes:2018emh,Kannike:2019wsn,Dev:2019njv,Kannike:2019mzk}.
Conformal symmetry also proposed as a possible solution for hierarchy problem \cite{Bardeen}.

Strongly first order electroweak phase transition can take place in the early Universe when two local minima of free energy (potential)
co-exist for some range of temperatures. It is a necessary condition in generating the observed baryon asymmetry in the universe and provides one of the three Sakharov conditions \cite{Sakharov:1967dj}, i.e., an out-of-equilibrium environment, in the framework of electroweak baryogenesis (for a recent review see \cite{Morrissey:2012db}). On the other hand, this violent phenomenon can lead to large anisotropic fluctuations in the energy-momentum tensor generating stochastic GW background \cite{Steinhardt:1981ct,Hogan:1984hx,Witten:1984rs,Hogan,Turner:1990rc,Kosowsky:2001xp}. This signal can potentially be probed in future space based GW detectors such as Laser Interferometer Space
Antenna (LISA) \cite{Seoane:2013qna,Audley:2017drz}, Big Bang Observer (BBO) \cite{Corbin:2005ny},
Deci-hertz Interferometer Gravitational wave Observatory
(DECIGO) \cite{Seto:2001qf}, and Ultimate-DECIGO (U-DECIGO) \cite{Kudoh:2005as}.

After two local minima of the free energy
co-exist at a critical temperature, the relevant scalar field can quantum mechanically tunnel into the new phase. This phenomenon continues via the nucleation of bubbles which expand and eventually collide with each other leaving a significant background of GWs.
According to simulations of GW backgrounds from cosmic phase transitions, there are three spectral contributions: 1) the collision spectrum which is the direct result of bubbles of true vacuum colliding \cite{Kosowsky:1991ua,Kosowsky:1992rz,Kosowsky:1992vn,Kamionkowski:1993fg,Caprini:2007xq,Huber:2008hg}, 2) sound wave spectrum which is the consequence of the fluid dynamics following such collisions, dominating in most relevant
scenarios \cite{Hindmarsh:2013xza,Giblin:2013kea,Giblin:2014qia,Hindmarsh:2015qta}, and 3) the turbulence spectrum, which is generally subdominant \cite{Caprini:2006jb,Kahniashvili:2008pf,Kahniashvili:2008pe,Kahniashvili:2009mf,Caprini:2009yp,Kisslinger:2015hua}.
GWs can prevail to the present times and perhaps be detected in space-based GW detectors.

Scale-invariant extensions of the SM can also provide a DM candidate (for recent papers see e.g. \cite{Okada:2012sg,Farzinnia:2013pga,
Wang:2015cda,Ghorbani:2015xvz,Ahriche:2015loa,Karam:2015jta,Karam:2016rsz,Khoze:2016zfi,Ahriche:2016ixu,Oda,YaserAyazi:2018lrv,Mohamadnejad:2019wqb,Jung:2019dog}).
Particle nature of the DM is another important puzzle in particle cosmology \cite{Bertone:2004pz}. There are in principle three ways to search for such exotic particle: 1) direct detection 2) indirect detection, and 3) collider searches.
Direct detection experiment such as the LUX \cite{Akerib:2016vxi}, PandaX-II \cite{Tan:2016zwf} and XENON1T \cite{Aprile:2018dbl} are gradually approaching the neutrino backgrounds which is usually considered as the ultimate sensitivity of future direct detection experiments \cite{Billard:2013qya}. Neutrino background (neutrino-floor) puts a limit on discovery potential of DM.
To this day, DM search experiments have not found any evidence.
However, null results of DM detection does not exclude the possibility of observing a GW signal from a dark sector. And in some models GW signals could  be a unique probe of the thermal DM paradigm.
In the absence of DM signal below neutrino-floor, GW experiments may serve as a new approach to probe the DM models.
Consequently, GW detectors can be a vital tools in exploring possibilities for DM models, complementing existing  efforts at colliders, direct and indirect detection experiments.

In this paper, we will study a conformal model \cite{YaserAyazi:2019caf} which is well motivated from a DM perspective or by naturalness arguments.
The model is a scale-invariant extension of the SM, enlarged by a dark $ U(1)_{D} $ gauge symmetry which provides a viable vector DM candidate. There are only two additional fields as well as free parameters in this model and the model can overcome constraints such as DM relic density, and direct and indirect detection upper bound limits.
For the parameter space constrained by DM relic density,
strongly first-order electroweak phase transition can also take place.
The main interest of the present paper is the GW signal produced during the electroweak phase transition. We use a benchmark point of the parameter space with below the neutrino-floor DM-nucleon cross section and show that the amplitude and frequency of phase transition GW spectrum fall within the observational window
of eLISA. Since, this particular choice of the parameter space is not constrained by colliders, and direct or indirect detection, therefore, GW signal plays an important role in probing the model for the chosen benchmark point.

This paper is structured as follows. In Sec.~\ref{sec2} we  introduce the model and review DM phenomenology. In this section, we choose a benchmark point below the neutrino-floor for the rest of the paper.
In Sec.~\ref{sec3}, we study effective potential. GW signal produced during first order electroweak phase transition as well as its discovery prospects
are presented in Sec.~\ref{sec4}, after which Sec.~\ref{sec5} comprises a summary and our conclusion.

\section{Review of the model and DM phenomenology \label{sec2}}

Let us first give an overview of the model presented in \cite{YaserAyazi:2019caf}.
The beyond SM fields content of the model are a complex scalar field ($ \phi $) and a vector field ($ V_{\mu} $). The model is a conformal extension of SM, enlarged by a dark Abelian gauge symmetry. The scalar field $ \phi $ has a unit charge under dark $ U(1)_D $ symmetry and $ V_{\mu} $ is the corresponding Abelian gauge field serving as DM particle.
These two fields are neutral under SM gauge group and SM fields are singlet under $ U(1)_D $.
There is also a discrete $ Z_{2} $ symmetry, under which SM particles are singlet and the vector field $ V_{\mu} $ and the scalar field $ \phi $ transform as follows:
\begin{equation}
V_{\mu} \rightarrow - V_{\mu} \, , \quad \phi \rightarrow \phi^{*}. \label{1}
\end{equation}
Note that, $ Z_{2} $ symmetry forbids the kinetic mixing between the vector field $ V_{\mu} $ and SM $ U_{Y}(1) $ gauge boson $ B_{\mu} $, therefore, the vector field $ V_{\mu} $ is stable and can be considered as a viable DM candidate.

The conformal model may be organized into three sectors:
1) the visible sector which consists of the SM fields without Higgs potential ($ {\cal L}_{VS} $), 2)
the dark sector which consists of the vector DM, $ V_{\mu} $, together with the scalar field $ \phi $ (${\cal L}_{DS}$), and 3) scale invariant tree-level potential ($V_{tree}$).
The Lagrangian is given by
\begin{equation}
 {\cal L} ={\cal L}_{VS} + {\cal L}_{DS} - V_{tree}, \label{2}
\end{equation}
where
\begin{equation}
{\cal L}_{DS} = (D_{\mu} \phi)^{*} (D^{\mu} \phi) - \frac{1}{4} V_{\mu \nu} V^{\mu \nu} , \label{3}
\end{equation}
with  $ D_{\mu} \phi = (\partial_{\mu} + i g_{v} V_{\mu}) \phi $ and $ V_{\mu \nu} = \partial_{\mu} V_{\nu} - \partial_{\nu} V_{\mu}  $. 
The most general scale-invariant potential which is renormalizable and invariant
under $ Z_{2} $ and gauge symmetry is
\begin{equation}
V_{tree} = \frac{\lambda_{H}}{6} (H^{\dagger}H)^{2} + \frac{\lambda_{\phi}}{6} (\phi^{*}\phi)^{2} + 2 \lambda_{\phi H} (\phi^{*}\phi) (H^{\dagger}H) , \label{4}
\end{equation}
where $ H $ is the Higgs doublet.
In Eq.~(\ref{4}), the third term is the only connection between the dark and the visible sector.

In unitary gauge, $ H^{\dagger} = \frac{1}{\sqrt{2}} (0 \quad h_{1}) $ and $ \phi = \frac{1}{\sqrt{2}} h_{2} $, therefore, tree-level potential is given by
\begin{equation}
V_{tree} = \frac{1}{4 !} \lambda_{H} h_{1}^{4} + \frac{1}{4 !} \lambda_{\phi} h_{2}^{4} + \frac{1}{2} \lambda_{\phi H} h_{1}^{2} h_{2}^{2}, \label{5}
\end{equation}
where $ h_{1,2} $ are real scalar fields.
Vacuum stability requires $ \lambda_{H,\phi} > 0 $ and $ \lambda_{\phi H} < 0 $,. Furthermore, non-zero VEV of $ h_{1,2} $ scalar fields demands $ \lambda_{H} \lambda_{\phi} = (3! \lambda_{\phi H})^{2} $.

\begin{table*}[t]
\centering
\begin{tabular}{c @{\hskip 0.4cm}c @{\hskip 0.4cm}c @{\hskip 0.4cm}c @{\hskip 0.4cm}c @{\hskip 0.4cm}c @{\hskip 0.4cm}c @{\hskip 0.4cm}c} 
\hline \\ [-2ex]
$ \sin\alpha $   &  $ g_{v} $   &   $ m_{\varphi} $(GeV)   &    $ \nu $(GeV)   &    $ m_{V} $(GeV) &   $ \Omega h^{2} $  & $ \sigma_{DM-N} $ (zb) & $\langle \sigma v \rangle$ ($ cm^{3}/s $) \\ 
[1ex] \hline \\ [-2ex]
$ 1.218\times10^{-1} $ & $ 5.653\times10^{-1} $ & $ 1.239\times10^2 $  & $ 2.019\times10^3 $ &  $ 1.133\times10^3 $  & $ 1.184\times10^{-1} $ &  $ 8.608\times10^{-4} $ &  $ 2.254\times10^{-26} $ \\
[1ex] \hline
\end{tabular}
\caption{A benchmark point of the model and corresponding DM relic density, SI DM-nucleon cross section, and DM total annihilation cross section. For this benchmark point, DM-nucleon cross section is below the neutrino-floor limit.}
\label{table1}
\end{table*}

The Local minimum of the two-variable potential (\ref{5}) defines a direction in field-space known as flat direction \cite{Gildener:1976ih}. Along this direction $ V_{tree} = 0 $, while in other directions $ V_{tree} > 0 $. Therefore, tree-level potential only vanishes along the flat direction where
\begin{equation}
\tan \alpha = \frac{h_{1}}{h_{2}} = \sqrt{- \frac{3! \lambda_{\phi H}}{\lambda_{H}}}. \label{6}
\end{equation}
Naturally, we expect higher-loop contributions be dominated along this direction and they should determine the local minimum of the full potential containing higher-loop effects.
Indeed, for some mass spectrum of the model, 1-loop effective potential, $ V_{eff}^{1-loop} $, gives a small curvature in the flat direction with $ V_{eff}^{1-loop} < 0 $. Therefore, considering 1-loop effect, the potential has a global minimum point in field space and consequently, symmetry breaking can take place. We assume $ \nu_{1} $ and $ \nu_{2} $ are VEVs of $ h_{1} $ and $ h_{2} $ where  $ \nu_{1} = 246 $ GeV. Now consider mass eigenstates $ h $ and $ \varphi $,
\begin{align}
& h =  cos \alpha \, h_1 - sin \alpha \, h_2 , \nonumber \\
& \varphi =  sin \alpha \, h_1 + cos \alpha \, h_2 , \label{7}
\end{align}
where $ \alpha $ is the angle between flat direction and $ h_{2} $ axis in field-space. Therefore, along the flat direction $ \left\langle h \right\rangle = 0 $, and all massive particles get mass when $ \left\langle \varphi \right\rangle \neq 0 $. In our formulation, $ h $ is perpendicular to the flat direction and we identify it as the SM-like Higgs observed at the LHC with $ m_{h} = 125 $ GeV.
At the classical tree-level, the scalon field $ \varphi $ is  massless, however, the 1-loop corrections give a mass to this field via Gildener-Weinberg mechanism \cite{Gildener:1976ih}. Regarding 1-loop effect, the scalon mass is given by
\begin{equation}
m_{\varphi}^{2} = \frac{1}{8 \pi^{2} \nu^{2}} \left( m_{h}^{4} + 6  m_{W}^{4} + 3  m_{Z}^{4} + 3  m_{V}^{4} - 12 m_{t}^{4}   \right) , \label{8}
\end{equation}
where $ m_{V,W,Z,t} $ being the masses for vector DM, W and Z gauge bosons, and top quark, respectively, and $ \nu = \sqrt{\nu_{1}^{2} + \nu_{2}^{2}} $ .
 
After symmetry breaking, all the four dimension-less parameters of the model, i.e., $ \lambda_{H,\phi H,\phi} $ and $g_{v}$, will be determined by DM mass $ m_{V} $, $ m_{h} $, $ \nu_{1} $, and $ \nu_{2} $.
Since we have already determined $ m_{h} $ and $ \nu_{1} $, therefore, the model has only two independent parameters. Here, we choose, DM mass $ m_{V} $ and $ \nu $ as our two-dimensional parameter space.
According to Eq. (\ref{8}), $ m_{\varphi} $ is also determined by this two-dimensional parameter space.

\begin{figure}[ht]
\centering
\includegraphics[scale=0.35]{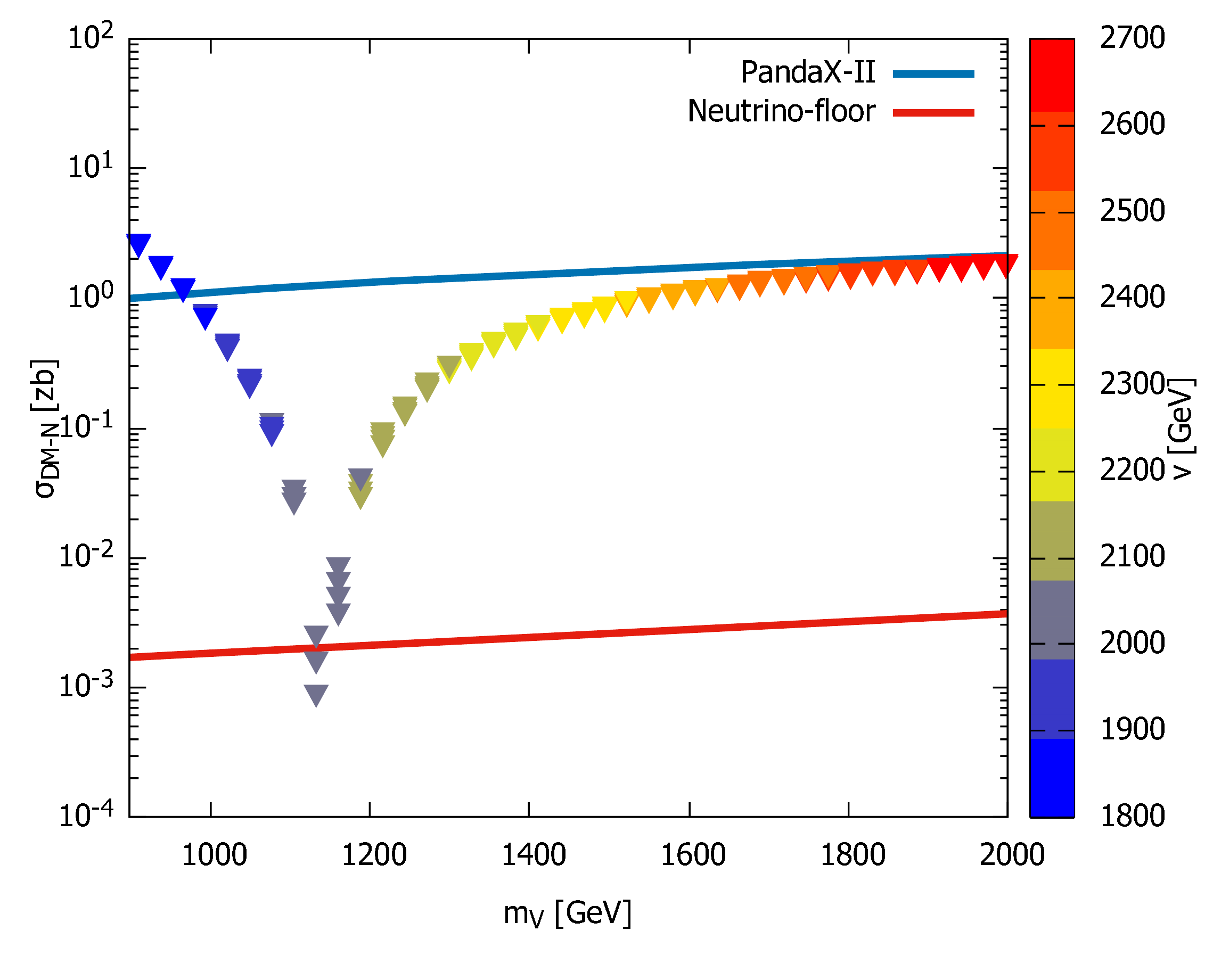}
\caption{\small \label{Direct} DM-nucleon cross section for a parameter spacce already constrained by DM relic density, $ \Omega h^{2} = 0.12 $ \cite{Aghanim:2018eyx}.} 
\end{figure}

We have recently studied the DM phenomenology of such a model \cite{YaserAyazi:2019caf}. Here we briefly report the main results.
In our model, for a narrow region of the parameter space, a direct detection of the vector DM is hopeless due to the neutrino-floor which represents an irreducible background, see FIG.~\ref{Direct}.

In this figure, spin-independent (SI) DM-nucleon cross section versus DM mass is depicted. PandaX-II \cite{Tan:2016zwf} upper bound as well as neutrino-floor \cite{Billard:2013qya} limit is shown for comparison. As it is seen, for a narrow region of parameter space, DM-nucleon cross section has a
dip below the neutrino-floor. The origin of this dip, is the proportionality of DM-nucleon cross section with $ (\frac{1}{m_{\varphi}^{2}} - \frac{1}{m_{h}^{2}})^{2} $ \cite{YaserAyazi:2019caf}. Therefore, for the parameter space around $ m_{\varphi} \simeq m_{h} $, we expect such a dip.
 
As we mentioned before, only a small portion
of the parameter space is below neutrino-floor. This region of parameter space is beyond the sensitivity of future Direct or indirect detection experiments \cite{YaserAyazi:2019caf}. However, we show that one can probe this blind-spot of the model using space-based GW detectors.
Since the region below the neutrino-floor is narrow, we choose only one benchmark for the rest of this paper, see TABLE~\ref{table1}. 

\begin{figure*}[t]
\begin{center}
\includegraphics[scale=0.45]{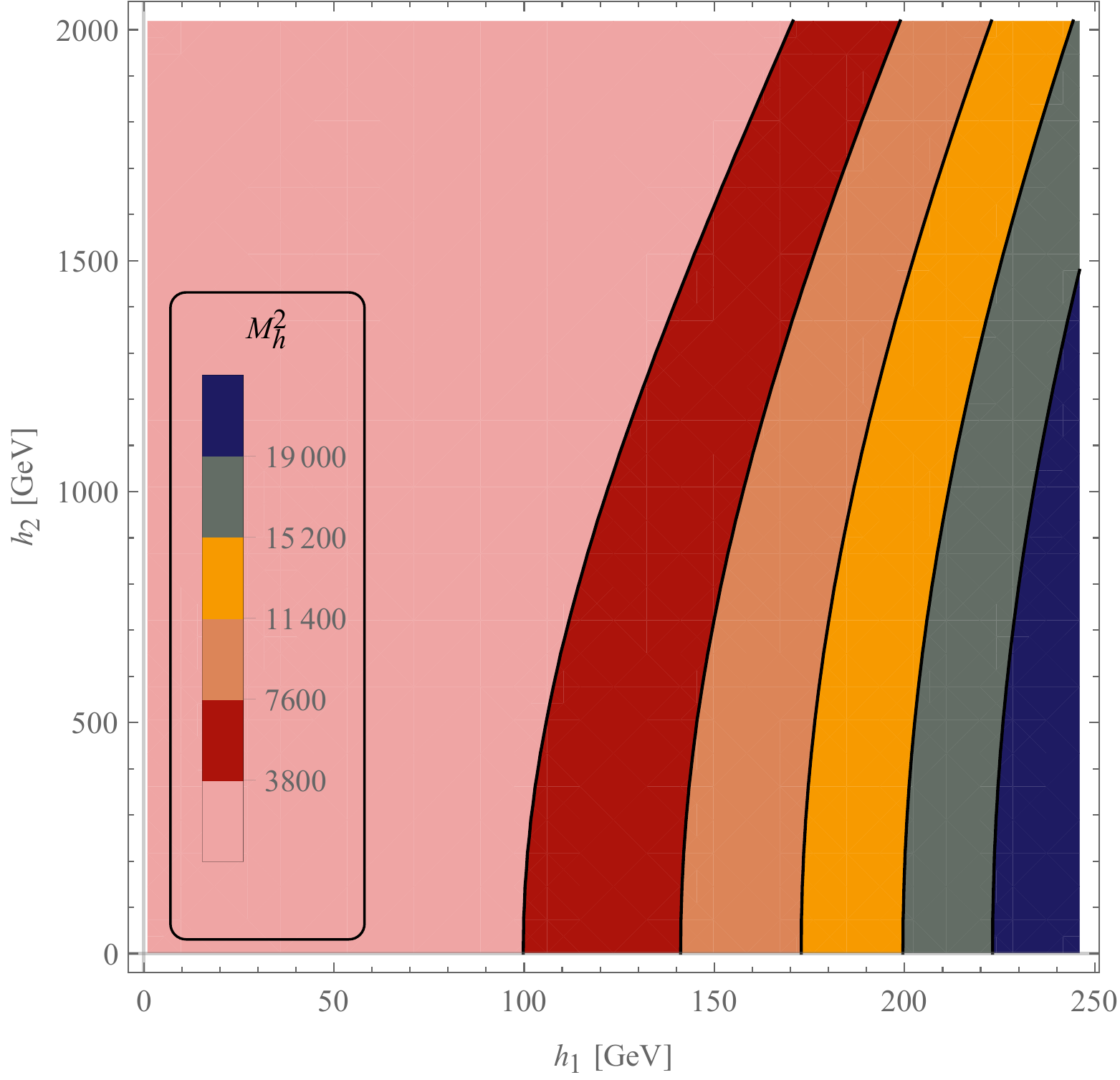}
\hspace{1.7cm}
\includegraphics[scale=0.45]{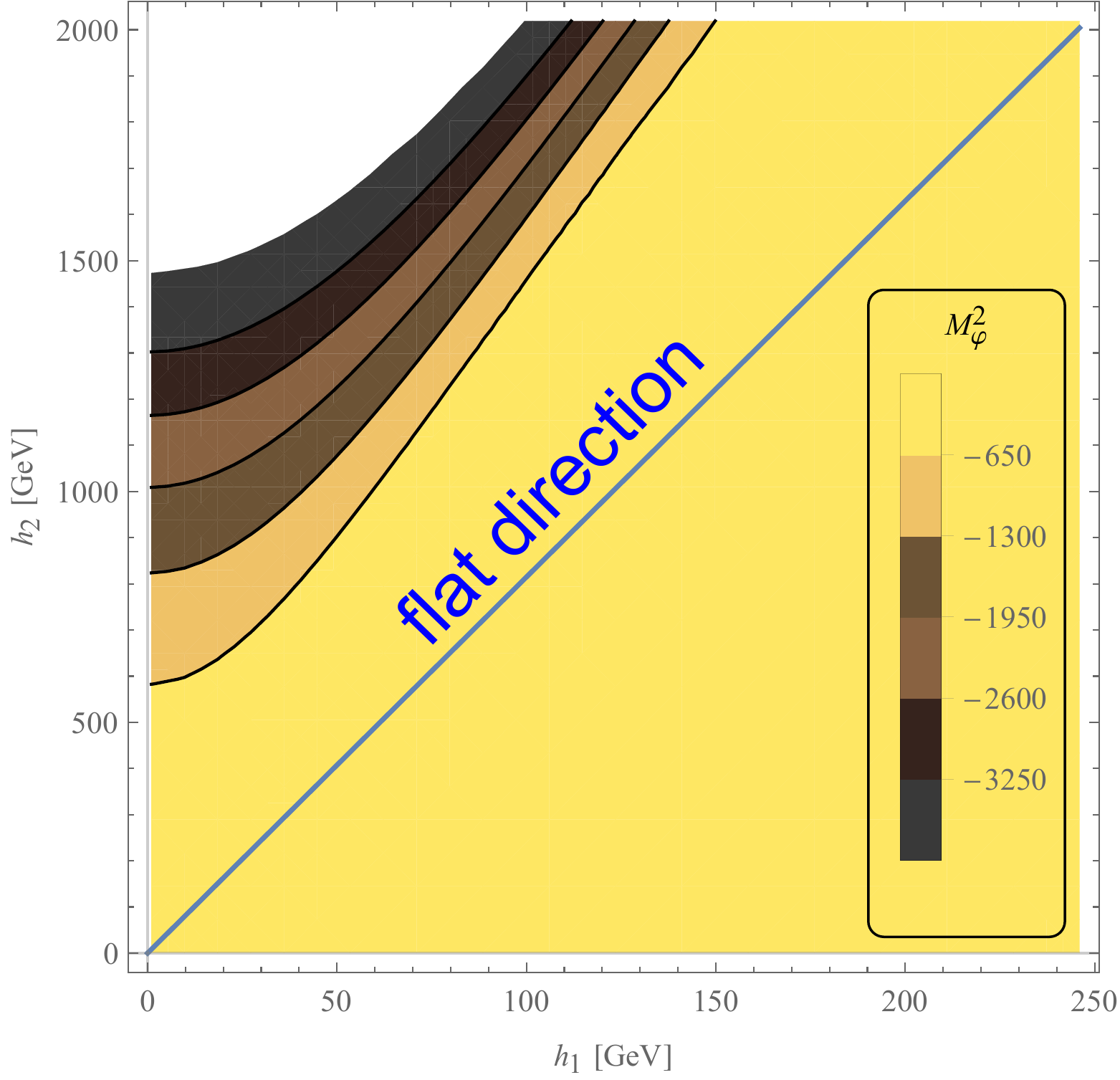}
\caption{\small \label{flat} Field-dependent masses of scalars. $ M_{h}^{2} $ (left) is positive, while $ M_{\varphi}^{2} $ (right) is negative, except along the flat direction where $ M_{\varphi}^{2} = 0 $. In the white region, $ M_{\varphi}^{2} $ will be complex.}
\end{center}
\end{figure*}

\section{Effective potential \label{sec3}}
The effective potential is composed of three parts: 1) classical or tree-level potential, 2) zero temperature 1-loop potential known as Coleman-Weinberg potential, and 3) 1-loop finite temperature potential. In the following we study these three pieces of full effective potential.
\subsection{Tree-level potential}
The tree-level potential is given in Eq.~(\ref{5}).
In the scalar sector, tree-level field-dependent masses correspond to the eigenvalues of the Hessian matrix:
\begin{align}
&{\cal{H}}_{ij} \equiv \frac{\partial^{2} V_{tree}}{\partial h_{i} \partial h_{j}} \nonumber \\
&\Rightarrow {\cal{H}}  =\begin{pmatrix} 
\frac{1}{2}\lambda _H h_1^2 +\lambda _{\text{$\phi H$}} h_2^2  & 2 \lambda _{\text{$\phi H$}}  h_1 h_2  \\
2 \lambda _{\text{$\phi H$}} h_1 h_2  & \frac{1}{2} \lambda _{\phi } h_2^2 + \lambda _{\text{$\phi H$}}  h_1^2 
\end{pmatrix}   , \label{9}
\end{align}
and are given by
\begin{align}
&M_{h,\varphi}^{2} = \frac{1}{4} \left( \left(\lambda _H+2 \lambda _{\text{$\phi H$}}\right) h_1^2+ \left(\lambda _{\phi }+2 \lambda _{\text{$\phi H$}}\right) h_2^2 \right) \pm \nonumber \\
& \sqrt{\frac{1}{16} \left( \left(\lambda _H-2 \lambda _{\text{$\phi H$}}\right) h_1^2 - \left(\lambda _{\phi }-2 \lambda _{\text{$\phi H$}}\right) h_2^2 \right){}^2+4 \lambda _{\text{$\phi H$}}^2 h_1^2 h_2^2 } . \label{10}
\end{align}
For the top quark and the gauge bosons the field dependent masses are
\begin{align}
&M_{t}^{2}  = \frac{\lambda_{t}^{2}}{2} h_{1}^{2}, \quad M_{V}^{2}  = g_{v}^{2} h_{2}^{2}, \nonumber \\
& M_{W}^{2}  = \frac{g^{2}}{4} h_{1}^{2},  \quad  M_{Z}^{2} = \frac{g^{2}+g'^{2}}{4} h_{1}^{2}, \label{11}
\end{align}
where $ \lambda_{t} $ denotes the top quark Yukawa coupling, and $ g_{v} $, $ g $, and $ g' $ are dark $ U(1)_{D} $, $ SU(2)_{L} $ and $ U(1)_{Y} $ gauge couplings, respectively.

According to Eqs.~(\ref{11}), it is obvious that $ M_{t}^{2} $, $ M_{V}^{2} $, $ M_{W}^{2} $, and $ M_{Z}^{2} $ are positive. 
As FIG.~\ref{flat} shows, $ M_{h}^{2} $ is also positive, however, $ M_{\varphi}^{2} < 0 $ except along the flat direction where $ M_{\varphi}^{2} = 0 $. Here, we exclude the field space with $ M_{\varphi}^{2} < 0 $ and only consider flat direction. 

\subsection{Coleman-Weinberg potential}
The Coleman-Weinberg potential is a sum of
1PI 1-loop diagrams with arbitrary numbers of external fields and particles running in the loop and it is given by \cite{Coleman:1973jx}
\begin{equation}
V^{1}_{\rm CW} =  \frac{1}{64 \pi^{2}}  \sum_{k=1}^{n} g_{k}  M_{k}^{4}  \left(   \ln \frac{M_{k}^{2}}{\Lambda^{2}} - C_{k} \right) , \label{12}
\end{equation}
where $ C_{k}=3/2 $ ($ 5/6 $) for scalars/spinors (vectors), $ M_{k} $ is the tree-level field-dependent mass of particle $ k $ given in Eqs.~(\ref{10}) and (\ref{11}), and $ g_{k} $ presents the number of degrees of freedom  given by
\begin{equation}
g_{k} =  (-1)^{2 s_{k}} q_{k} N_{k} (2 s_{k} + 1), \label{13}
\end{equation}
where $ s_{k} $ is the spin, $ N_{k} $ the number of colors and $ q_{k}=1 $ (2) for neutral (charged) particles.

In Eq.~(\ref{12}), to get a real Coleman-Weinberg potential, the field dependent mass squared of particles can not be negative. Thus the allowed field space should be along the flat direction.
On the other hand, along this direction $ M_{\varphi} =0 $, and we do not consider scalon field contribution in Coleman-Weinberg potential (\ref{12}). Moreover, the Goldstone bosons are massless along the flat direction and they do not contribute in the minimum of the tree-level potential. Therefore, we do not consider field dependent masses of Goldstone bosons in Eq.~\eqref{12} as well.

Note that, along the flat direction, for five remained field dependent mass contributions, we can substitute $ M_{k} \rightarrow \frac{m_{k}}{\nu} \varphi $, where $ m_{k} $ is the measured mass of particle $ k $.
Regarding this substitution in Eq.~(\ref{12}) results the well-known Gildener-Weinberg formula \cite{Gildener:1976ih}
\begin{equation}
V^{1}_{\rm GW} = A \varphi^{4} + B \varphi^{4} \ln \frac{\varphi^{2}}{\Lambda^{2}} , \label{14}
\end{equation}
where
\begin{align}
& A =  \frac{1}{64 \pi^{2} \nu^{4}}  \sum_{k=1}^{n} g_{k}  m_{k}^{4} \left(  \ln \frac{m_{k}^{2}}{\nu^{2}} - C_{k}  \right)   , \nonumber \\
& B = \frac{1}{64 \pi^{2} \nu^{4}} \sum_{k=1}^{n} g_{k}  m_{k}^{4} . \label{15}
\end{align}

As we mentioned before, along the flat direction $ V_{tree}=0 $. Therefore, to find the true vacuum, one should find the minimum of the 1-loop potential (\ref{14}), given by
\begin{equation}
\langle \varphi \rangle = \nu = \Lambda e^{-(\frac{A}{2B} + \frac{1}{4})} . \label{16}
\end{equation}
Combining Eq.~(\ref{14}) and Eq.~(\ref{16}) we can substitute RG scale $ \Lambda $ and find a simple expression for the 1-loop potential in terms of the true vacuum expectation value $ \nu $ and $ B $ coefficient:
\begin{equation}
V^{1}_{\rm GW} = B \varphi^{4} \left(  \ln \frac{\varphi^{2}}{\nu^{2}} - \frac{1}{2} \right) . \label{17}
\end{equation}
From the above equation, one can find $ m_{\varphi}^{2} = \frac{d^2 V^{1}_{\rm GW}}{d \varphi^{2}} \bigg\rvert_{\nu} $, see Eq.~(\ref{8}).
Although, the $ \varphi $ scalar obtains a radiatively generated mass at 1-loop level, Goldstone bosons remain massless to all orders in perturbation theory.

\subsection{Finite temperature potential}
\begin{figure}[ht]
\centering
\includegraphics[scale=0.5]{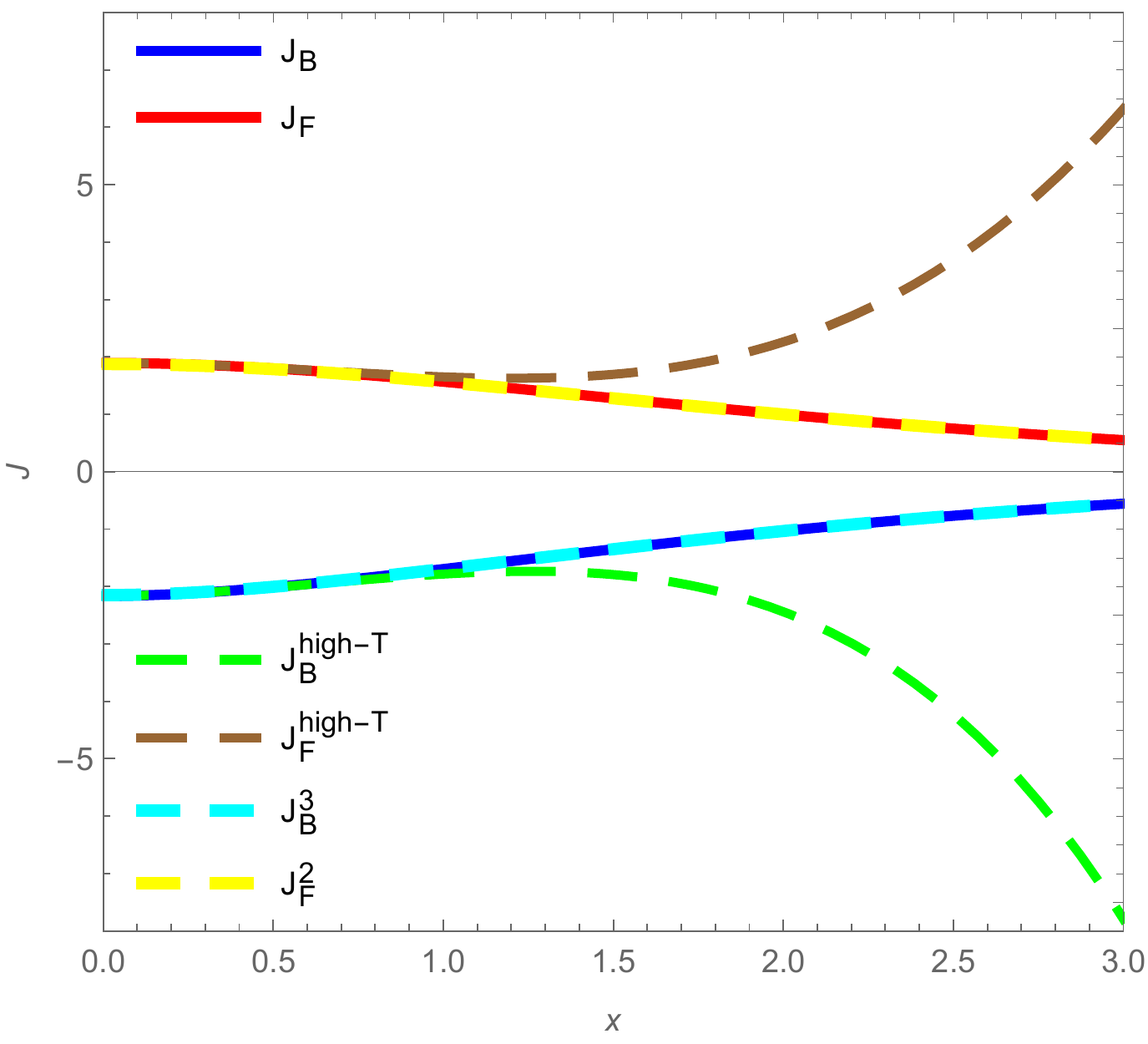}
\caption{\small \label{J} Thermal functions and their different approximations. } 
\end{figure}

Now we study finite-temperature 1-loop effective potential which enables us to compute scalar field vacuum
expectation values, in the background of a thermal bath with temperature $ T $.
The 1-loop finite-temperature corrections are given by \cite{Dolan:1973qd}
\begin{equation}
V_{T}^{1} = \frac{T^{4}}{2 \pi^{2}}  \sum_{k=1}^{n} g_{k} J_{\text{B,F}} (\frac{M_{k}}{T}) , \label{18}
\end{equation}
with thermal functions
\begin{equation}
J_{\text{B,F}}(x) =  \int_{0}^{\infty} dy \, y^{2} \ln \left(1 \mp e^{- \sqrt{y^{2}+x^{2}}} \right),   \label{19}
\end{equation}
vanishing as $ T \rightarrow 0 $.
Although these integrals cannot be expressed in terms of standard functions, their
numerical evaluation is rather straightforward and one can approximate them in different limits.
For example, in the high temperature limit ($ x \ll 1 $), $ J_{\text{B}}(x) $ and $ J_{\text{F}}(x) $ have very useful closed forms (see appendix C in \cite{Dolan:1973qd}),
\begin{align}
&J_{\text{B}}^{\text{high-T}}(x) = -\frac{\pi ^4}{45}  +\frac{\pi ^2 }{12} x^2  -\frac{\pi}{6}  x^3 -\frac{1}{32} x^4 \ln \left(\frac{x^2}{a_{b}}\right) , \nonumber \\
&J_{\text{F}}^{\text{high-T}}(x) = \frac{7 \pi ^4}{360} -\frac{\pi ^2}{24} x^{2} -\frac{1}{32} x^4 \ln \left(\frac{x^{2}}{a_{f}}\right) , \label{20}
\end{align}
where $ a_b = \pi ^2 \exp \left(\frac{3}{2}-2 \gamma_{E} \right) $, $ a_f = 16 \, a_b $ and $ \gamma_{E} $ denotes the Euler-Mascheroni constant.
The thermal functions $ J_{\text{B}}(x) $ and $ J_{\text{F}}(x) $ also have a useful expansion
in terms of modified Bessel functions of the second kind (see Appendix \ref{Appendix})
\begin{align}
&J_{\text{B}}^{m}(x) = -\sum _{k=1}^m \frac{1}{k^2} x^2 K_2\left(k x\right), \nonumber \\
&J_{\text{F}}^{m}(x) = -\sum _{k=1}^m \frac{(-1)^k }{k^2} x^2 K_2\left(k x\right). \label{21}
\end{align}
The above sum representations are convergent as $ m \rightarrow \infty $ and in this limit $ J^{m}(x) \rightarrow J(x) $.
In FIG.~\ref{J}, we have depicted $ J(x) $, $ J^{\text{high-T}}(x) $, and  $ J^{m}(x) $. 
As it is seen in this figure, with the log term included, high-T expansion for the thermal functions is accurate to better than 10 percent even
for $ x \sim \, (1-1.5) $ (depending on the function), but breaks down completely beyond that.
However, the summation in Eq. (\ref{21}) can be truncated at a few terms, for example m = 2 for $ J_{F} $ and m = 3 for $ J_{B} $, and still yield a very good accuracy (see FIG.~\ref{J}).

In order to consider as well the resummation of
the Matsubara zero modes, it is essential to
resum the thermal masses by substituting $ M^{2}  \rightarrow M_{tree}^{2}  +\Pi  $.
In the
standard method, $ \Pi $ is taken to be the leading contribution in temperature to the 1-loop thermal mass \cite{Carrington:1991hz}. For scalars  $ \Pi $
can be estimated by differentiating $ V_{T}^{1} $ with respect to $ \varphi $ yielding $ \Pi \propto T^2 $.
This replacing automatically contains daisy contributions to all orders in the effective potential. 

Using high-T approximation, the field dependent terms in logs cancel between  $ V_{T}^{1}  $ and  $ V_{\rm CW}^{1} $. The $ x^{2} $ term gives an overall contribution proportional to $ T^{2} \Pi_{i} $. If we use only the leading-order contribution to $ \Pi_{i} $ in temperature, $ T^{2} \Pi_{i} $ will be field-independent.
Therefore, only $ x^{3} $ term is left, which can be entrap by adding $ V_{ring} $. This means, effective potential contains four terms:
\begin{equation}
V_{eff} = V_{tree} + V^{1}_{\rm CW}  + V_{T}^{1}  + V_{ring} , \label{22}
\end{equation}
where $ V_{ring} $ is daisy term \cite{Carrington:1991hz}:
\begin{equation}
V_{ring}  = \sum_{k=1}^{n} \frac{g_{k} T}{12 \pi} \left(  M_{k}^{3} - (M_{k}^{2}  + \Pi_{k}(T))^{3/2} \right) .  \label{23}
\end{equation}
Including $ V_{ring} $ amounts to resumming the IR-divergent terms to the Matsubara zero mode propagator. It is equivalent to substitute $ M^{2}  \rightarrow M_{tree}^{2} +\Pi (T) $  in the
full effective potential, assuming that only the
thermal mass of the zero mode is relevant, which means using high-temperature approximation.

The sum in Eq.~(\ref{23}) runs only over longitudinal degrees of freedom of the gauge
bosons and scalars. The thermal
masses of the gauge bosons are given by \cite{Carrington:1991hz}
\begin{align}
&\Pi_{W} = \left( \frac{5}{6}+\frac{n_{f}}{3}\right)  g^{2} T^2, \quad  \Pi_{V} = \frac{2}{3} g_{v}^{2} T^2, \nonumber \\
&\Pi_{Z/\gamma}=
\begin{pmatrix} 
\left( \frac{5}{6}+\frac{n_{f}}{3}\right)  g^{2} & 0 \\
0 & \left( \frac{1}{6}+\frac{5n_{f}}{9}\right)  g'^{2} 
\end{pmatrix}
T^2, \nonumber \\
\label{24}
\end{align}
where $ n_{f}=3 $ is the number of fermionic generations.
For scalars we have
\begin{align}
&\Pi_{\text{Scalar}} = \begin{pmatrix} 
\frac{\partial^{2} V^{high-T}}{\partial h_{1}^{2}} & \frac{\partial^{2} V^{high-T}}{\partial h_{1} \partial h_{2}} \\
\frac{\partial^{2} V^{high-T}}{\partial h_{2} \partial h_{1}} & \frac{\partial^{2} V^{high-T}}{\partial h_{2}^2}
\end{pmatrix}
 \nonumber \\
&=\begin{pmatrix} 
\frac{\lambda_{H}}{24} + \frac{\lambda_{\phi H}}{12}  + \frac{3g^2}{16}  + \frac{g'^2}{16} + \frac{\lambda_{t}^2}{4}  & 0 \\
0 & \frac{\lambda_{\phi}}{24} + \frac{\lambda_{\phi H}}{12} + \frac{g_{v}^2}{4} 
\end{pmatrix} T^2 , \label{25}
\end{align}
where $ V^{high-T} $ is derived from Eq.~(\ref{18}) using high temperature approximation~(\ref{20}) with only $ x^2 $ term, i.e.,
\begin{equation}
V^{high-T}  = \frac{T^{2}}{24} \left(  \sum_{B} g_{B} M_{B}^{2} - \frac{1}{2} \sum_{F} g_{F} M_{F}^{2}   \right)  . \label{26}
\end{equation}
The factors
of the scalar couplings in Eq.~(\ref{25}) are different from
the results in the literature (e.g. ref.~\cite{Carrington:1991hz}) since we did
not include the impact of the Goldstone bosons (as the same as ref.~\cite{Prokopec:2018tnq}).

Considering Eq.~(\ref{22}) along the flat direction (where $ V_{tree}=0 $), we can obtain a one-dimensional effective potential, $ V_{eff}(\varphi,T) $ which contains Gildener-Weinberg potential (\ref{17}) and thermal contributions (\ref{18}) and (\ref{23}).
In the next section using this potential, we study the phase
transition and GW signal.

\section{Electroweak phase transition and GW signal \label{sec4}}
\begin{figure}[ht]
\begin{center}
\includegraphics[scale=0.5]{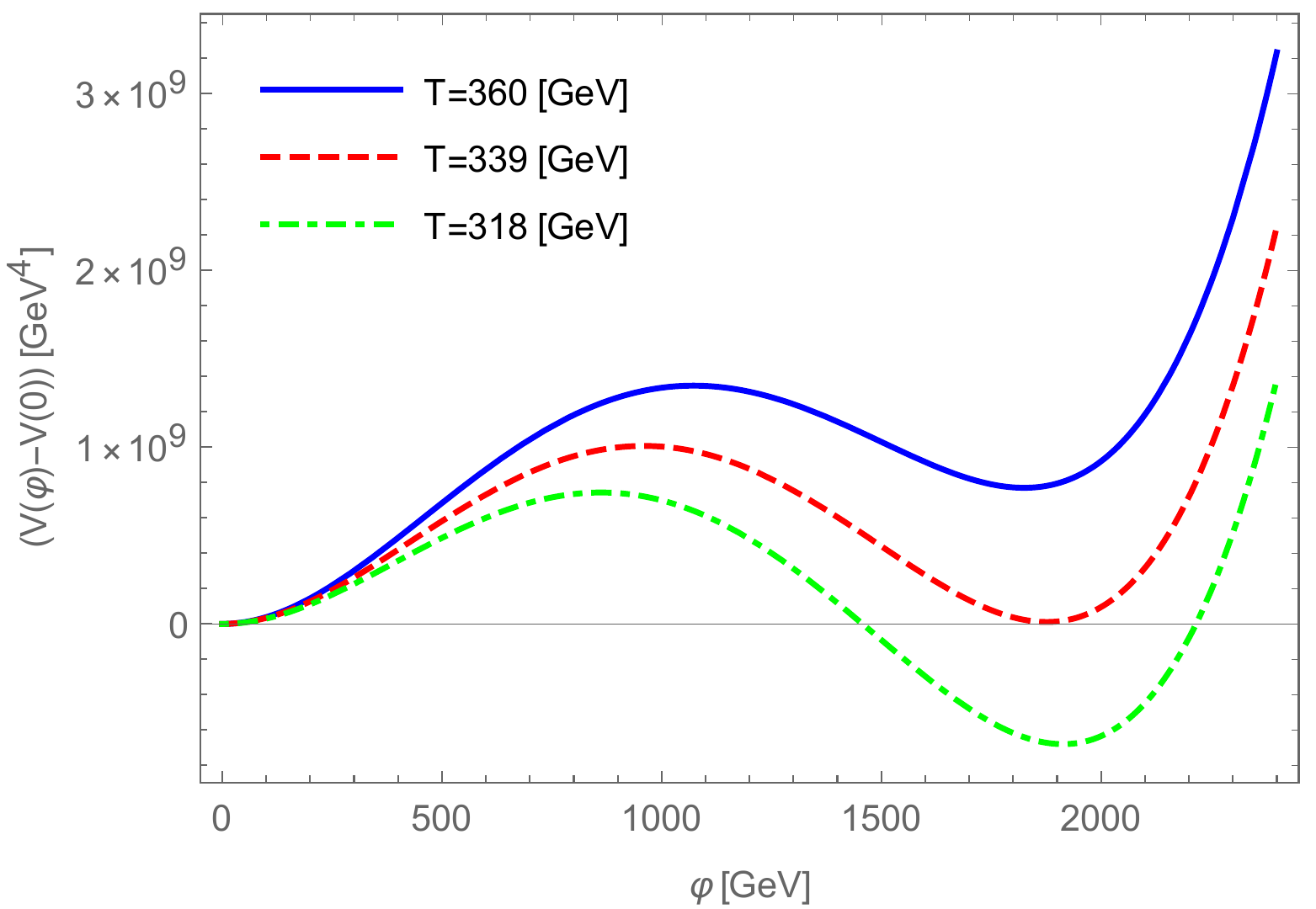}
\caption{\small \label{V} The effective potential
which accounts for the full 1-loop thermal effects including the re-summed
daisy diagrams plotted for three different temperatures.
At the critical temperature $ T_c=339 $ GeV the minima at $ \varphi=0 $ and $ \varphi \simeq 1900 $ GeV are degenerate.
} 
\end{center}
\end{figure}

The electroweak phase transition takes place after the temperature of the universe drops below the critical temperature (see FIG.~\ref{V}) and the minimum with non-zero scalon VEV becomes the global minimum. For the parameters in TABLE \ref{table1}, the critical temperature is $ T_c=339 $  GeV at which the effective potential has two degenerate minimums.

After the minimum with non-zero scalon VEV becomes the global minimum, thermal fluctuations eventually excite the field enough to cross the potential barrier.
For conformal extensions of SM, the potential barrier disappears only for $ T=0 $, thus
the electroweak phase transition in scale-invariant
models is always of first order for any finite temperature.

The decisive quantity
is the temperature at which phase transition proceeds via nucleation
and consequent expansion of bubbles inside of which the field
is in the broken phase of the model. The rate of bubble nucleation per unit of time and volume is given by \cite{Linde:1981zj}
\begin{equation}
\Gamma (T) \simeq T^4 \left( \frac{S_3(T)}{2 \pi T} \right)^{3/2} e^{-S_3(T) / T} , \label{27}
\end{equation}
where
\begin{equation}
S_3(T) =  4 \pi \int_{0}^{\infty} dr \, r^2 \left( \frac{1}{2} \left( \frac{d \varphi}{dr} \right)^2 + V_{eff}(\varphi,T)  \right) ,  \label{28}
\end{equation}
is the three-dimensional Euclidean action for a spherical
symmetric bubble. The differential equation
\begin{equation}
\frac{d^2 \varphi}{dr^2} + \frac{2}{r} \frac{d \varphi}{dr}  = \frac{d  V_{eff}(\varphi,T) }{d \varphi} ,   \label{29}
\end{equation}
minimizes $ S_3 $, therefore, the field $ \varphi $ obtained by solving the equation (\ref{29}) has the largest contribution to the bubble nucleation rate.
The solution of (\ref{29}) starts for $ r=0 $ close to a specified true vacuum in field space with $ \frac{d \varphi}{dr} =0 $ and asymptotically come near to a specified false
vacuum as $ r \rightarrow \infty $. 
The nucleation temperature $ T_n $ is defined as the temperature
at which the probability of one bubble nucleation per horizon volume in Hubble time approaches unity:
\begin{equation}
\frac{4 \pi}{3} \frac{\Gamma(T_{n})}{H(T_{n})^{4}} \simeq 1   . \label{30}
\end{equation}
The Hubble parameter as a function of temperature is given by \cite{Ellis:2018mja}
\begin{equation}
H(T) = \frac{\pi T^2}{3 M_{pl}} \sqrt{\frac{g_{*}}{10}} , \label{31}
\end{equation}
where $ M_{pl} \sim 10^{19} $ GeV and $ g_{*} \sim 100 $ are the Planck mass and the effective number of relativistic degrees of freedom in the thermal plasma, respectively. One can also estimates $ T_n $ by the condition $ S_3(T_n)/T_n \simeq 140 $ \cite{Apreda:2001us}.

\begin{figure}[ht]
\begin{center}
\includegraphics[scale=0.5]{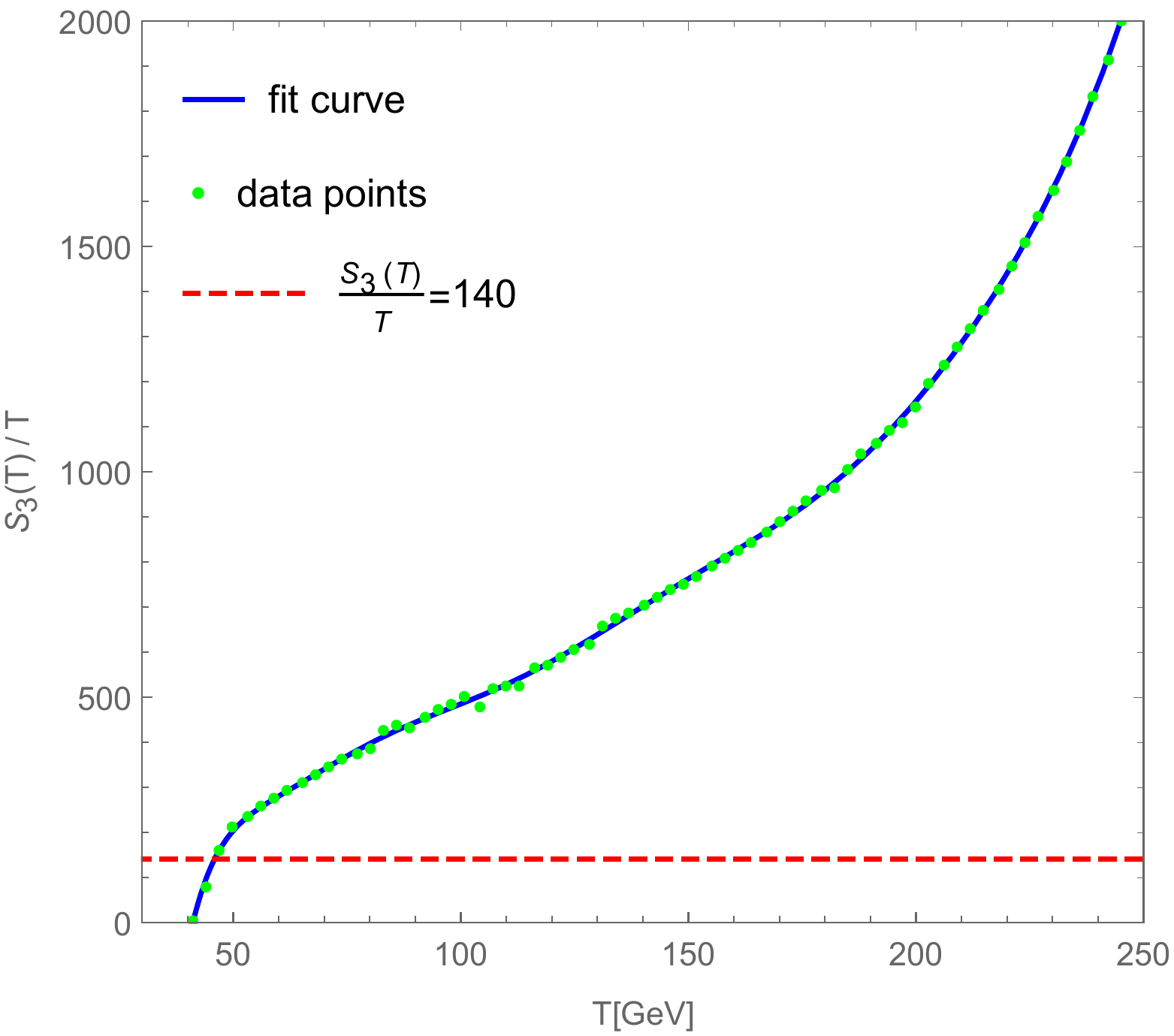}
\caption{\small \label{S3} The blue solid line presents
$ S_3/T $ versus $ T $, and the dashed horizontal red line shows $ S_3/T = 140 $ where nucleation occurs. The green dots are obtained using {\tt AnyBubble} package.} 
\end{center}
\end{figure}

In order to solve Eq.~(\ref{29}) and find the Euclidean action  (\ref{28}), we have used {\tt AnyBubble} package \cite{Masoumi:2017trx}. The result is depicted in FIG.~\ref{S3}.
Using condition (\ref{30}), we determine the nucleation temperature, $ T_n =47 $ GeV, which is much lower than the critical temperature.
Therefore, phase transition proceeds after
a large amount of supercooling
and transition should be very strong, consequently we expect a fairly large GW signal produced at the temperature $ T_{*} $.
For typical phase transitions with negligible reheating, $ T_{*} $ is approximately equivalent to the nucleation temperature $ T_n $.

Now we study
stochastic GW background produced by strong first-order electroweak phase
transitions. The resulting contributions come from three processes: 
\begin{itemize}
\item collisions of bubble walls and shocks in the plasma,
\item sound waves to the stochastic background after collision of bubbles but before expansion has
dissipated the kinetic energy in the plasma, and
\item turbulence forming after the bubbles have
collided. 
\end{itemize}
These three processes can coexist, and each one contributes to the stochastic GW background:
\begin{equation}
h^{2} \Omega_{\rm tot} \simeq h^{2} \Omega_{\rm coll} + h^{2} \Omega_{\rm sw} + h^{2} \Omega_{\rm turb} . \label{32}
\end{equation}

All of these three contributions are controlled by four thermal parameters (see TABLE~\ref{table2}):
\begin{itemize}
\item $ T_n $ : the nucleation temperature,
\item  $ \alpha $ : the ratio of the free energy density difference between the true and false vacuum and
the total energy density,
\begin{equation}
\alpha= \frac{\Delta \left( V_{eff} - T \frac{\partial V_{eff}}{\partial T} \right)  \bigg \rvert_{T_n}}{\rho_{*}}  ,  \label{33}
\end{equation}
where $ \rho_{*} $ is given by
\begin{equation}
\rho _*=\frac{\pi ^2 g_*}{30} T_n^4 , \label{34}
\end{equation}
\item $ \beta $ : the inverse time duration of the phase transition,
\begin{equation}
\frac{\beta}{H_*}= T_n \frac{d}{d T} \left(  \frac{S_3 (T)}{T}\right) \bigg \rvert_{T_n} , \label{35}
\end{equation}
\item $ v_w $ : the velocity of the bubble wall
which is anticipated to be close to 1 for the strong transitions \cite{Bodeker:2009qy}.
\end{itemize}

\begin{table}[t]
\centering
\begin{tabular}{c @{\hskip 0.4cm}c @{\hskip 0.4cm}c @{\hskip 0.4cm}c} 
\hline \\ [-2ex]
$ T_c  $(GeV)   &  $ T_n $(GeV)   &   $ \alpha $   &    $ \beta/H_* $   \\ 
[1ex] \hline \\ [-2ex]
$ 339 $ & $ 47 $ & $ 24 $  & $ 808 $ \\
[1ex] \hline
\end{tabular}
\caption{Parameters of the phase transition for the benchmark point in TABLE~\ref{table1}.}
\label{table2}
\end{table}

The electroweak phase transition proceeds by the nucleation and expansion of bubbles
of the new phase. Note that isolated spherical bubbles can not be a source of GWs and these waves
arise during the collision of the bubbles.
The collision contribution to the spectrum is given by \cite{Huber:2008hg}
\begin{align}
h^2\Omega_{\rm coll}(f) =& 1.67 \times 10^{-5}   \, \left( \frac{\beta}{H_*} \right)^{-2} \left( \frac{\kappa \alpha}{1+\alpha} \right)^2  
  \nonumber\\ &\times \left( \frac{g_*}{100} \right)^{-\frac{1}{3}} \left(\frac{0.11\,v_w^3}{0.42+v_w^2}\right) \, S_{\rm coll} , \label{36}
\end{align}
where $ S_{\rm coll} $ parametrises the spectral shape and is given by
\begin{equation}
S_{\rm coll}=  \frac{3.8 \left(f/f_{\text{coll}}\right)^{2.8}}{2.8 \left(f/f_{\text{coll}}\right)^{3.8}+1} ,
\label{37}
\end{equation}
where
\begin{align}
f_{\text{coll}}=& 1.65 \times 10^{-5} \left( \frac{ 0.62 }{ v_w^2-0.1 v_w+1.8} \right) \nonumber \\
& \times  \left( \frac{\beta}{H_*} \right) \left( \frac{ T_n}{100} \right) \left( \frac{g_*}{100} \right)^{1/6} \, \text{Hz} .
\label{38}
\end{align}

Bubble collision produces bulk motion in the fluid in the form of sound waves which generates GWs. This is the dominant
contribution to the GW signal which is given by \cite{Hindmarsh:2015qta}
 \begin{align}
h^2\Omega_{\rm sw}(f) =& 2.65 \times 10^{-6}   \, \left( \frac{\beta}{H_*} \right)^{-1} \left( \frac{\kappa_{v} \alpha}{1+\alpha} \right)^2  
  \nonumber\\ &\times \left( \frac{g_*}{100} \right)^{-\frac{1}{3}} v_{w} \, S_{\rm sw}. \label{39}
\end{align}
The spectral shape of $ S_{\rm sw} $ is
\begin{equation}
S_{\rm sw}=  \left(f/f_{\text{sw}}\right)^3 \left(\frac{7}{3  \left(f/f_{\text{sw}}\right)^2+4}\right)^{3.5} ,
\label{40}
\end{equation}
where
\begin{align}
f_{\text{sw}}= 1.9 \times 10^{-5} \frac{1}{v_w}
 \left( \frac{\beta}{H_*} \right) \left( \frac{ T_n}{100} \right) \left( \frac{g_*}{100} \right)^{1/6} \, \text{Hz} .
\label{41}
\end{align}

Bubble collisions can also generate turbulence in the plasma which its contribution to the GW spectrum is given by \cite{Caprini:2009yp}
 \begin{align}
h^2\Omega_{\rm turb}(f) =& 3.35 \times 10^{-4}   \, \left( \frac{\beta}{H_*} \right)^{-1} \left( \frac{\kappa_{\rm turb} \alpha}{1+\alpha} \right)^{3/2}  
  \nonumber\\ &\times \left( \frac{g_*}{100} \right)^{-\frac{1}{3}} v_{w} \, S_{\rm turb} , \label{42}
\end{align}
where
\begin{equation}
S_{\rm turb}=  \frac{\left(f/f_{\text{turb}}\right)^3}{\left(1+8 \pi  f/h_*\right) \left(1+f/f_{\text{turb}}\right){}^{11/3}} ,
\label{43}
\end{equation}
and
\begin{align}
f_{\text{turb}}= 2.27 \times 10^{-5} \frac{1}{v_w}
 \left( \frac{\beta}{H_*} \right) \left( \frac{ T_n}{100} \right) \left( \frac{g_*}{100} \right)^{1/6} \, \text{Hz} .
\label{44}
\end{align}
In Eq.~(\ref{43}), $ h_* $ is the value of the inverse Hubble time at GW production, redshifted to today,
\begin{equation}
 h_* = 1.65 \times 10^{-5}  \left( \frac{ T_n}{100} \right) \left( \frac{g_*}{100} \right)^{1/6} .
 \label{45}
\end{equation} 

\begin{figure}[ht]
\begin{center}
\includegraphics[scale=0.5]{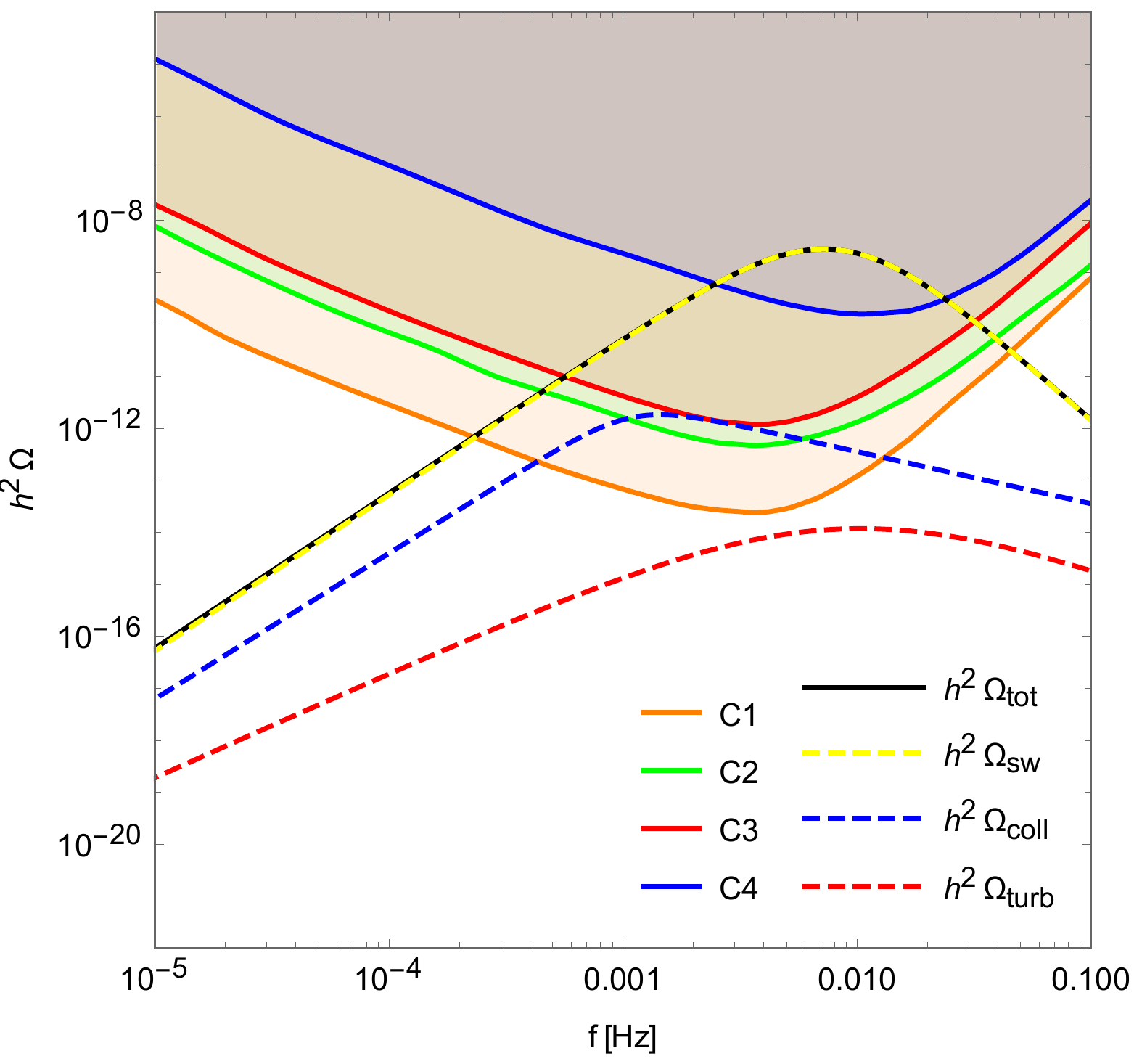}
\caption{\small \label{GW} 
GW spectra for the phase transition parameters (see TABLE~\ref{table2}).
The black line denotes the total GW spectrum, the yellow line the contribution from sound
waves, the blue line the contribution from bubble collisions, and the red line the contribution from turbulence.
Sensitivity curves of the four configurations of eLISA detector (C1-C4) are also depicted \cite{Caprini:2015zlo}.
The amplitude and peak frequency of phase transition GW spectrum fall within the observational window of eLISA.
} 
\end{center}
\end{figure}

In computing GW spectrum we have used \cite{Caprini:2015zlo,Kamionkowski:1993fg}
\begin{align}
& \kappa = \frac{1}{1+0.715 \, \alpha} (0.715 \, \alpha +\frac{4}{27} \sqrt{\frac{3 \alpha }{2}}) , \nonumber \\
&  \kappa_{v} = \frac{\alpha}{0.73 +0.083 \sqrt{\alpha }+\alpha} , \quad \kappa_{\rm turb} = 0.05 \kappa_{v},
 \label{46}
\end{align}
where $ \kappa $, $ \kappa_{v} $, and $ \kappa_{\rm turb} $ denote the fraction of latent heat that is transformed into gradient energy of the Higgs-like field, bulk
motion of the fluid, and MHD turbulence, respectively.
The result is depicted in FIG.~\ref{GW}.
As it is seen in this figure, the peak frequency of GW is about 0.005 Hz and it is strong enough to be detected by eLISA detector.

\section{Conclusion \label{sec5}}
Dark matter is believed to be five times as predominant as visible matter and its gravitational impacts are seen throughout the Universe. However, we have yet to see GWs caused by dark matter, and we can think of numerous ways this might happen.
Indeed, detecting such signals can constrain dark matter models.
Currently, the connection between DM and GWs is largely unexplored, and much remains to be done
to make existing constraints better and to study the prospects for identifying DM using upcoming experiments.
In this paper, this connection is studied through electrowaek phase transition. The GW spectrum from a first order phase transition
is assumed to follow a broken power-law. However, the Standard Model does not feature a first-order phase transition, therefore, observed GW background of this kind would point uniquely to new
physics.

Detecting GWs of cosmological origin is not an easy task. 
There is an enormous foreground due to astrophysical sources which in principle makes detection challenging. Once the signals from merging neutron stars and stellar mass black holes have been recognized and
removed, the main sources of foreground signals
are galactic and extragalactic binaries. The galactic background generated by binary stars in the Milky Way is many
times larger in amplitude than both the extragalactic foreground and eLISA’s design sensitivity. Furthermore, the galactic background, being mostly concentrated in the galactic plane, can be removed duo to its anisotropy.
Here we study a stochastic background of GWs, which is statistically homogeneous and isotopic. Therefore,
the real problem is irreducible background coming from extragalactic binary stars which is dominated
by emission from white dwarf binaries at a level of approximately 
$ h^2 \Omega  = 10^{-12} $ \cite{Farmer:2003pa}. These signals typically appear in millihertz regime \cite{Bender:1997hs,Evans:1987qa} and other signals of GW in this regime may be hidden by them. Nevertheless, there are some techniques that may detect stochastic GWs background in the presence of both instrument noise and foregrounds of GW spectrum, see, e.g., \cite{Adams:2013qma}.

In this paper, we have studied the GW signatures associated with
the strong electroweak phase transition in the early universe in a scale-invariant vector DM model.
In this model, both DM mass and the Higgs potential originate from the effects of a real scalar field (scalon) whose mass term dynamically develop through the spontaneous breaking of classical scale-invariant symmetry. To study phase transition, we obtained effective potential including three terms: 1) tree-level potential, 2) Coleman-Weinberg 1-loop potential, and 3) finite temperature potential with daisy diagrams contributions. Since, we have considered the flat direction in the field space, the effective potential only contains the Coleman-Weinberg and thermal terms.
We show that the nucleation temperature is much lower than the
critical temperature, i.e.,
the phase transition proceeds after large super-cooling.
To compute the GW spectrum, we have considered a benchmark point of the parameter space below the neutrino-floor and see that the amplitude and frequency of GW fall within the observational window
of eLISA.

Although, our DM model cannot be confirmed only by detecting GWs, it certainly can be disapproved with not detecting such signals. 
Indeed, detecting GWs alone barely reveals the particle physics model behind the phase transition. Hence, GWs can be a vigorous tools in exploring possibilities for DM models, complementing already existing efforts at colliders, direct and indirect detection experiments. For the particular choice of the parameter space of this paper, these efforts cannot constrain the model at the present. However, with developing these experiments, our model can be distinguished from other models with the same GWs signals containing DM candidates or without them. For example, in this model, the scalon modifies the Higgs potential and, although its direct detection at the LHC is hopeless due to the small signal to background ratio, it could be possible at future collider \cite{Curtin:2014jma}. Moreover, indirect collider searches, e.g., coming from the modification of the triple Higgs coupling or the Zh production at lepton colliders, provide another probe of the model.
Therefore, both collider and GW signals can provide a realistic and complementary test of the model. However, in the absence of stronger colliders, the planned GW experiments can play an important role in investigating our model, and possibly other DM models with strong electroweak phase transition, especially in probing below the neutrino-floor.

\appendix
\section{Expansion of thermal functions} \label{Appendix}
The natural logarithm has Maclaurin series
\begin{equation}
\ln (1 \mp x) = - \sum_{k=1}^{\infty} \frac{(\pm 1)^{k}}{k}  x^{k} . \label{A1}
\end{equation}
Using the above formula in the thermal functions,
\begin{equation}
J_{\text{B,F}}(x) =  \int_{0}^{\infty} dy \, y^{2} \ln \left(1 \mp e^{- \sqrt{y^{2}+x^{2}}} \right),   \label{A2}
\end{equation}
we get
\begin{align}
J_{\text{B,F}}(x) &=  \int_{0}^{\infty} dy \, y^{2}   (  - \sum_{k=1}^{\infty} \frac{(\pm 1)^{k}}{k}  e^{- k \sqrt{y^{2}+x^{2}}}  ) \nonumber \\
&=  - \sum_{k=1}^{\infty} \frac{(\pm 1)^{k}}{k} \int_{0}^{\infty} dy \, y^{2}     e^{- k \sqrt{y^{2}+x^{2}}} . \label{A3}
\end{align}
If we make the substitutions
\begin{align}
y^{2}+x^{2} &= t^2 x^2, \nonumber \\
\Rightarrow dy \, y^{2}  &= dt \, t \, x^3 (t^2 - 1)^{1/2} , \label{A4}
\end{align}
we obtain
\begin{equation}
J_{\text{B,F}}(x) =  - \sum_{k=1}^{\infty} \frac{(\pm 1)^{k}}{k} x^3 \int_{1}^{\infty} dt \, t  (t^2 - 1)^{1/2}     e^{- k xt} . \label{A5}
\end{equation}
Now let
\begin{align}
& u =  e^{- k xt} , \nonumber \\
 &v =  \frac{1}{3} (t^2 - 1)^{3/2} \Rightarrow dv = dt \, t (t^2 - 1)^{1/2} , \label{A6}
\end{align}
therefore,
\begin{equation}
J_{\text{B,F}}(x) =  - \sum_{k=1}^{\infty} (\pm 1)^{k} \frac{x^4}{3} \int_{1}^{\infty} dt  (t^2 - 1)^{3/2}     e^{- k xt} . \label{A7}
\end{equation}
In order to derive the above formula, the integration by parts is performed. Equation (\ref{A7}) can be written as
\begin{align}
J_{\text{B,F}}(x) &=  - \sum_{k=1}^{\infty}  \frac{(\pm 1)^{k}}{k^{2}} x^{2} \frac{4}{3} (\frac{kx}{2})^2 \int_{1}^{\infty} dt  (t^2 - 1)^{3/2} e^{- k xt} \nonumber \\
&=  - \sum_{k=1}^{\infty}  \frac{(\pm 1)^{k}}{k^{2}} x^{2} K_{2} (kx), \label{A8}
\end{align}
where we have used the integral representation of modified Bessel functions of the second kind \cite{arfken}
\begin{equation}
K_{\nu}(x)=\frac{\sqrt{\pi }}{\Gamma \left(\nu +\frac{1}{2}\right)} \left(\frac{x}{2}\right)^{\nu }  \int_1^{\infty } \, dt \left(t^2-1\right)^{\nu -\frac{1}{2}} e^{-x t} . \label{A9}
\end{equation}
Equation~(\ref{A8}) is the same as Eq.~(\ref{21}) with 
 $ m \rightarrow \infty $.

\providecommand{\href}[2]{#2}
\end{document}